\documentclass[twocolumn,showpacs,showkeys]{revtex4}
\usepackage{graphicx}
\usepackage{bm}
\usepackage{color}
\usepackage{amsmath}
\usepackage{natbib}

\begin{document}

\title{On a mechanism of high-temperature superconductivity: Spin-electron acoustic wave as a mechanism for the Cooper pair formation.}

\author{Pavel A. Andreev}
\email{andreevpa@physics.msu.ru}
\author{P. A. Polyakov}
\email{pa.polyakov@physics.msu.ru}
\author{L. S. Kuz'menkov}%
\email{lsk@phys.msu.ru}
\affiliation{Faculty of physics, Lomonosov Moscow State University, Moscow, Russian Federation.}

\date{\today}

\begin{abstract}
We have found the mechanism of the electron Cooper pair formation via the electron interaction by means of the spin-electron acoustic waves. This mechanism takes place in metals with rather high spin polarization, like ferromagnetic, ferrimagnetic and antiferromagnetic materials. The spin-electron acoustic wave mechanism leads to transition temperatures 100 times higher than the transition temperature allowed by the electron-phonon interaction. Therefore, spin-electron acoustic waves give the explanation for the high-temperature superconductivity. We find that the transition temperature has strong dependence on the electron concentration and the spin polarization of the electrons.
\end{abstract}


\maketitle





The high-temperature superconductivity phenomenon was discovered in 1986 \cite{Bednorz ZfP B 86}. It was found in metallic, oxygen-deficient compounds in the Ba-La-Cu-O systems and, later, in other similar structures (see for instance \cite{Wu PRL 87}), which are antiferromagnetic materials. In the conventional superconductivity the phase transition temperatures is below 20 K, at normal pressure. The high-temperature superconductivity appears at temperatures up to $T_{c}=160$ K \cite{Schilling Nat 93}, \cite{Nunez-Regueiro Science 93}. Hence, the area of the superconductivity is shifted from the liquid helium temperatures (4 K) to the liquid nitrogen temperatures (77 K).

Models for the conventional superconductivity were developed by Bardeen, Cooper, and Schrieffer \cite{Bardeen PR 57}, \cite{Bardeen PR 57 (2)} (BCS model) and Bogoliubov \cite{Bogoliubov JETP 58}. The mechanism of the conventional superconductivity is the formation of pairs of electrons with opposite spins and momentums (the Cooper pairs). This formation occurs via the electron-phonon interaction. The BCS model gives the following value for the energy gap in the spectrum of the elementary excitations
\begin{equation}\label{HTSC Tc}\Delta=2\hbar\omega_{D}e^{-2\pi^{2}\hbar^{3}/(|g|m_{e}p_{Fe})},\end{equation}
which is proportional to the temperature of the phase transition in the superconductive state $\Delta=1.75k_{B}T_{c}$. We see that the phase transition temperature is proportional to the Debye frequency of phonons $\omega_{D}=u_{ph}(6\pi^{2}n)^{\frac{1}{3}}$, where $u_{ph}$ is the phonon speed, $n$ is the electron concentration, $\hbar$ is the reduced Planck constant, $k_{B}$ is the Boltzmann constant, $m_{e}$ is the mass of electron, $p_{Fe}=(3\pi n)^{\frac{1}{3}}\hbar$ is the Fermi momentum, $g$ is the constant of the electron-phonon interaction arising in the Hamiltonian $H_{e-ph}=(g/V)\sum_{\textbf{p},\textbf{p}'}\hat{a}_{\textbf{p}',+}^{+}\hat{a}_{-\textbf{p}',-}^{+}\hat{a}_{-\textbf{p},-}\hat{a}_{\textbf{p},+}$, with $\hat{a}_{\textbf{p},s}^{+}$ and $\hat{a}_{\textbf{p},s}$ are the creation and annihilation operators of the electron in the quantum state with the momentum $\textbf{p}$ and the spin projection $s=\{+,-\}$ describing the spin-up and spin-down states, $V$ is the volume element (see Refs. \cite{Bardeen Sc 73}, \cite{Schrieffer RMP 99} for the further discussion of this model).

The contribution of the ion lattice into the electron-electron interaction and its relevance for the superconductivity were suggested by Frohlich \cite{Frohlich PR 50} and developed by Cooper \cite{Cooper PR 56}. Quasi particles different from phonons have been suggested as a mechanism of the electron-electron interaction (see for instance \cite{Schuht JLTP 83}). Discussion of the "superconductivity without phonons" was presented in Ref. \cite{Monthoux Nat 07}.

The model of the conventional superconductivity does not allow calculation of the electron-phonon interaction constant $g$. Comparison of the formula (\ref{HTSC Tc}) with the experimental data leads us to the conclusion that the exponent should be equal to 0.01 if we want to have transition temperatures $T_{c}$ below 20 K.

Bardeen, Cooper, and Schrieffer described a physical mechanism of the conventional superconductivity \cite{Bardeen PR 57}. However, there is an open problem to find a mechanism for the high temperature superconductivity. In this paper we suggest a mechanism of this phenomenon. We find that the high temperature superconductivity is related to the Cooper pair formation, but these pairs appear due to the interaction different from the phonon-electron interaction that accounts for  the conventional superconductivity. We discover the interaction between electrons via the spin-electron acoustic waves existing in the materials with partially spin polarized electrons \cite{Andreev PRE 15}. This mechanism can be described in terms of the BCS model with the extended Hamiltonian. As the result of application of the BCS model to the electron-spelnon (a quantum of the spin-electron acoustic wave) interaction we find phase transition temperatures about 100 K.

Two types of matter waves can exist in normal metals (we do not consider the propagation of electromagnetic radiation through the materials): the ion-acoustic and the Langmuir waves. The ion-acoustic wave (phonon) has linear dispersion dependence at small wave vectors $k$: $\omega=ku_{ph}$. Partially spin polarized metals, such as materials with the magnetic order, reveal special kind of longitudinal waves, which are called the spin-electron acoustic waves (SEAWs), with linear spectrum at small wave vectors \cite{Andreev PRE 15}. Below we apply the fluid model of dielectric permittivity of the medium to show the possibility of the Cooper pair formation. Therefore, the electron-spelnon interaction mechanism of the Cooper pair formation is suggested as a mechanism of the high-temperature superconductivity.

\emph{Conditions of SEAW appearance.} The equation of state for the pressure of spin-up $P_{\uparrow}$ and spin-down $P_{\downarrow}$ degenerate electrons appears as $P_{s}=\frac{(6\pi^{2})^{2/3}}{5}\frac{\hbar^{2}}{m}n_{s}^{5/3}$. Pressures of spin-up electrons and spin-down electrons differ due to different populations of the spin-up and spin-down degenerate electrons $n_{\uparrow}\neq n_{\downarrow}$. The pressure $P_{s}$ is related to the case of a single particle with a chosen spin direction occupying specified quantum state. As a consequence, the coefficient in the equation of state is $2^{2/3}$ times bigger than in the Fermi pressure. The difference of the spin-up and spin-down electrons' concentrations $\Delta n=n_{\uparrow}-n_{\downarrow}$ can be caused by the field of exchange interaction with the electrons in ion cores, forming the ferrimagnetic or antiferromagnetic states, or by the external magnetic field. To describe these effects we apply the separate spin evolution quantum hydrodynamics, which deals spin-up and spin-down electrons as two different fluids and shows the formation of the spin-electron acoustic waves \cite{Andreev PRE 15}.

\emph{Justification of the Cooper pair formation be means of SEAWs.} The electron-electron interaction in vacuum is described by the bare Coulomb potential $U_{Cr}=e^{2}/r$ with the following Fourier image $U_{Ck}=4\pi e^{2}/k^{2}$. In a medium this interaction is screened by other electrons and positively charged ions. When the frequencies considered are close to the eigen-frequencies of ion or electron motion then the resonance phenomena can occur. An overscreening can take place and two negative charges can attract each other. We can see the possibility of this phenomenon when considering the dielectric function $\varepsilon=\varepsilon(\omega, k)$, which contributes to the effective potential $U_{Ck}=4\pi e^{2}/(k^{2}\varepsilon)$.

In the regime of the electron-electron interaction via the acoustic phonon we can find the dielectric permittivity of the medium $\varepsilon_{IA}$ and represent it as follows
\begin{equation}\label{HTSC reverce varepsilon IA} \frac{1}{\varepsilon_{IA}}=\frac{k^{2}}{k_{S}^{2}+k^{2}}\biggl(1+\frac{\omega_{IA}^{2}}{\omega^{2}-\omega_{IA}^{2}}\biggr),\end{equation}
where
\begin{equation}\label{HTSC omega IA} \omega_{IA}^{2}(k)=\omega_{Li}^{2}/(1+k_{S}^{2}/k^{2})\simeq(m_{e}/m_{i})v_{Fe}^{2}k^{2}/3,\end{equation}
is the spectrum of the ion-acoustic waves, with $\omega_{Li}^{2}=4\pi e^{2}n/m_{i}$ is the ion Langmuir frequency, $k_{S}^{2}=3\omega_{Le}^{2}/v_{Fe}^{2}$, and $v_{Fe}=(3\pi^{2}n)^{\frac{1}{3}}\hbar/m_{e}$ is the Fermi velocity of the degenerate electrons, $m_{i}$ is the mass of ion, where the approximate expression is found in the small wave vector limit.

In the regime of partial spin polarization of the conducting electrons, with regard for the contribution of the SEAWs to the electron-electron interaction,  in accordance with Ref. \cite{Andreev PRE 15} (the contribution of the Coulomb exchange interaction in the dielectric function was found in Ref. \cite{Andreev 1504}), we obtain
\begin{equation}\label{HTSC reverce varepsilon SEAW} \frac{1}{\varepsilon_{SEAW}}=\frac{k^{2}}{k_{EA}^{2}+k^{2}}\biggl(1+\frac{\omega_{SEAW}^{2}}{\omega^{2}-\omega_{SEAW}^{2}}\biggr),\end{equation}
where
$$\omega_{SEAW}^{2}(k)=\omega_{Lu}^{2}/(1+k_{EA}^{2}/k^{2})$$
\begin{equation}\label{HTSC omega SEAW} \simeq(n_{u}/n_{d})(6\pi^{2}n_{d})^{\frac{2}{3}}\hbar^{2}k^{2}/3m_{e}^{2},\end{equation}
is the spectrum of the SEAWs, with $\omega_{Ls}^{2}$ $=4\pi e^{2}n_{s}/m_{e}$, are the partial Langmuir frequencies of the spin-up and spin-down electrons, $k_{EA}^{2}$$=\omega_{Ld}^{2}/\tilde{v}_{Fd}^{2}$, $n_{s}=\{n_{u},n_{d}\}$. Spectrum (\ref{HTSC omega SEAW}) also contains the modified Fermi velocity for the spin-down electrons containing the exchange Coulomb interaction \cite{Andreev 1504}: $\tilde{v}_{Fd}^{2}=(6\pi^{2}n_{d})^{\frac{2}{3}}\hbar^{2}/3m_{e}^{2}-\chi e^{2}n_{d}^{\frac{1}{3}}/m_{e}$, with $\chi=2^{\frac{4}{3}}\sqrt[3]{\frac{3}{\pi}}(1-\frac{(1-\eta)^{4/3}}{(1+\eta)^{4/3}})$, where the first term is one third of the square of the Fermi velocity of the spin-down electrons, and the second term is the exchange Coulomb interaction of spin-down electrons being in quantum states occupied by single electron.

Substituting formulae (\ref{HTSC reverce varepsilon IA}) and (\ref{HTSC reverce varepsilon SEAW}) into the effective Coulomb potential $U_{Ck}=4\pi e^{2}/(k^{2}\varepsilon)$ we find that each of them yields two types of contributions. In both cases the first term in the effective Coulomb potentials describes the screened Coulomb repulsion. The second term in the phonon regime, presented by formula (\ref{HTSC reverce varepsilon IA}), reflects the electron-electron interaction specified by phonons. This part of interaction shows the attraction between electrons at $\omega<\omega_{IA}$. Similarly, the second term in the reverse dielectric permittivity (\ref{HTSC reverce varepsilon SEAW}), describing the spelnon regime, gives the interaction specified by spelnons, which reveals the attraction between electrons at $\omega<\omega_{SEAW}$. Therefore, we can conclude that the electron-spelnon interaction provides a mechanism for the Cooper pair formation in the partially spin polarized materials.

We apply the BCS model to the extended Hamiltonian $H$ describing our partially spin polarized system. This Hamiltonian is composed of the electron $H_{e}$, phonon $H_{ph}$, and spelnon $H_{sp}$ Hamiltonians together with the interaction of the quasiparticles: the electron-phonon $H_{e-ph}$, electron-spelnon $H_{e-sp}$,  and phonon-spelnon $H_{ph-sp}$ interactions. Hamiltonians $H_{e}$, $H_{ph}$, $H_{e-ph}$ have the traditional form, for example the electron-phonon interaction Hamiltonian is presented above. In this paper we consider partially spin polarized conductivity electrons. Therefore, we have the energy of electron contribution in Cooper pairs equal to $E\leq E_{Fu}$, where we assume $n_{u}<n_{d}$, with the particle concentration of the spin-up $n_{u}$ and spin-down $n_{d}$ electrons, $E_{Fu}$ is the Fermi energy of the spin-up electrons. The focus of our analysis turns to the electron-spelnon interaction, hence we present this part of Hamiltonian $H_{e-sp}=(\varrho/V)\sum_{\textbf{p},\textbf{p}'}\hat{a}_{\textbf{p}',+}^{+}\hat{a}_{-\textbf{p}',-}^{+}\hat{a}_{-\textbf{p},-}\hat{a}_{\textbf{p},+}$, where $\varrho$ is the constant of the interaction. In accordance with the analysis presented above $H_{e-sp}$ describes an attractive interaction $\varrho <0$. As a result of the application of the BCS model we find the transition temperature
\begin{equation}\label{HTSC Tc for HTsc}T_{sp}=\frac{1.14\hbar}{k_{B}}\varpi e^{-2\pi^{2}\hbar^{3}/(|\varrho|m_{e}p_{Fu})},\end{equation}
where we have the Debye frequency of the spelnons $\varpi=u_{sp}(6\pi^{2}n_{u})^{\frac{1}{3}}$, with $u_{sp}$ is the spelnon speed resulting from the dispersion dependence (\ref{HTSC omega SEAW}).

\begin{figure}
\includegraphics[width=8cm,angle=0]{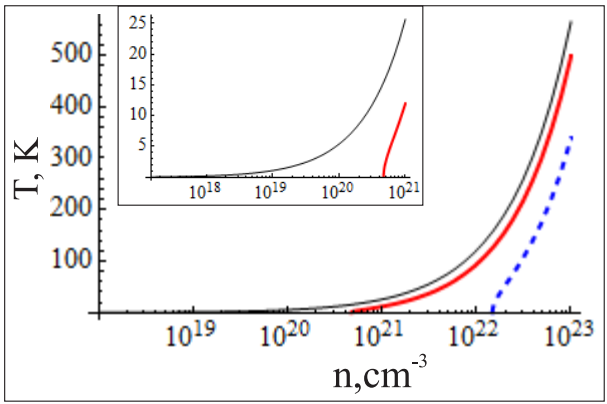}
\caption{\label{HTSC Tc on spin polarization} (Color online) The figure shows the transition temperature as a function of the electron concentration $n$ at different values of the spin polarization of the conductivity electrons in accordance with formula (\ref{HTSC Tc for HTsc}). The thin continuous (black) line describes regime $\eta=0.001$. Thick continuous (red) line corresponds to $\eta=0.015$. The dashed line is obtained for $\eta=0.05$. The small figure shows the area of small concentrations with more details.}
\end{figure}

At small wave vectors both the ion-acoustic waves and the SEAWs have linear spectrum giving us the "speeds of sounds". For the ion-acoustic wave we have $u_{ph}=\sqrt{m_{e}/3m_{i}}v_{Fe}$ and for SEAWs we find $u_{sp}=\omega_{SEAW}/k=\sqrt{n_{u}/3n_{d}}\tilde{v}_{Fd}\approx\sqrt{n_{u}/3n_{d}}(6\pi^{2}n_{d})^{\frac{1}{3}}\hbar/m_{e}$, where the approximate expression is obtained by dropping of the exchange part of the Coulomb interaction. The speed of sound defines the Debye frequency, which is proportional to the phase transition temperature  in the superconductive state. Therefore, we see that the SEAW mechanism of the Cooper pair formation gives us the transition temperature $T_{sp}\sim100\div200$ K, which is $u_{sp}/u_{ph}=2^{\frac{1}{3}}\sqrt{(n_{u}/n_{d})(m_{i}/m_{e})}\sim 10^{2}$ times larger than the transition temperature for the ion-acoustic mechanism that accounts for the conventional superconductivity. This is an estimate for low spin polarization of the electrons $\eta\sim0.001$. More accurate results for the transition temperature are presented in Fig. \ref{HTSC Tc on spin polarization}.

The electron-spelnon interaction constant $\varrho$ is an undefined quantity in the extended BCS model. However, we need its numerical value to estimate the transition temperature given by formula (\ref{HTSC Tc for HTsc}). At low spin polarization $\eta\ll 1$ the exponent in formula (\ref{HTSC Tc for HTsc}) is equal to the exponent in formula (\ref{HTSC Tc}) with the replacement of $g$ by $\varrho$. Assuming that the strength of the electron-spelnon interaction is approximately equal to the strength of the electron-phonon interaction we can take the value of the exponent from the electron-phonon mechanism of the Cooper pair formation and substitute it in our model. Therefore, we assume that difference between two models is in the prefactor before the exponents in formulae (\ref{HTSC Tc}) and (\ref{HTSC Tc for HTsc}). Under these assumptions we can calculate the transition temperature (\ref{HTSC Tc for HTsc}) and present our results at Fig. \ref{HTSC Tc on spin polarization}.

Our estimations give the transition temperature above 1 K at electron concentrations $n$ above $10^{18}$ cm$^{-3}$. These values appear at rather low spin polarization $\eta\sim 0.001$. For such spin polarization and larger concentrations $n\sim10^{21}$ cm$^{-3}$ we find larger transition temperatures. For instance, at $\eta=0.015$ and $n\simeq10^{22}$ cm$^{-3}$ we have $T_{c}=100$ K. Further increase in the concentration allows to achieve the transition to the superconductive state at room temperatures $T_{c}\sim300$ K, for instance, we find $n=2\times 10^{22}$ cm $^{-3}$ at $\eta=0.001$ or $n=9\times 10^{22}$ cm $^{-3}$ at $\eta=0.05$. Increasing the spin polarization (in area below $\eta=0.05$) of the conductivity electrons we decrease the temperature of transition in the superconductive state at fixed electron concentrations. We also find relation between temperature and the concentration at fixed spin polarization: $T_{c}^{3}/n^{2}=c$, where $c$ is a constant depending on the spin polarization $\eta$. $c$ monotonically decreases with the increase of the spin polarization $\eta$. At fixed electron concentration we find the decrease of the transition temperature with the increase of the spin polarization.

In the conventional superconductivity the isotopic effect takes place$T_{c}\sim m_{i}^{-0.5}$ since $T_{c}\sim u_{ph}\sim m_{i}^{-0.5}$, which has been experimentally observed by Maxwell \cite{Maxwell PR 50}. While in the regime of the electron-spelnon interaction, giving the high-temperature superconductivity, we find $T_{c}'\sim u_{sp}$ and the phase transition temperature does not demonstrate any isotopic effect. However it does depend on the spin polarization of the electrons.

In conclusion we should point out that the electron-spelnon interaction provides realistic mechanism for the superconductivity at high temperatures via the Cooper pair formation. This mechanism allows us toexpect the superconductivity at room temperatures if the electrons' concentration would be high enough.

In acknowledgements P.A. thanks the Dynasty foundation for financial support.

\end{document}